\documentstyle[prc,aps]{revtex}

\def\x{\mbox{\boldmath $x$}}
\def\z{\mbox{\boldmath $z$}}
\def\s{\mbox{\boldmath $s$}}
\def\h{\mbox{\boldmath $h$}}
\def\S{\mbox{\boldmath $S$}}
\def\N{\mbox{\boldmath $N$}}
\def\L{\mbox{\boldmath $L$}}

\def\p{\mbox{\boldmath $p$}}
\def\P{\mbox{\boldmath $P$}}
\def\q{\mbox{\boldmath $q$}}

\def\e{\mbox{\boldmath $e$}}
\def\N{\mbox{\boldmath $N$}}

\def\ss{\mbox{\boldmath $\sigma$}}

\begin{document}
\draft
\title{Proton recoil polarization in exclusive ($\e,\e'\p\p$) reactions}
\author{C.~Giusti and F.~D.~Pacati}
\address{Dipartimento di Fisica Nucleare e Teorica dell'Universit\`a, 
Pavia\\
and Istituto Nazionale di Fisica Nucleare, Sezione di Pavia, Italy}
\date{\today}
\maketitle

%%%%%%%%%%%%%%%%%%%%%%%%%%%%%%%%%%%%%%%%%%
\begin{abstract}
The general formalism of nucleon recoil polarization in the 
(${\vec e},e'{\vec N}N$) reaction is given. Numerical predictions are presented
for the components of the outgoing proton polarization and of the polarization
transfer coefficient in the specific case of the exclusive 
$^{16}$O(${\vec e},e'{\vec p}p$)$^{14}$C knockout reaction leading to discrete 
states in the residual nucleus. Reaction calculations are performed in a direct
knockout framework where final-state interactions and one-body and two-body
currents are included. The two-nucleon overlap integrals are obtained from a
calculation of the two-proton spectral function of $^{16}$O where long-range
and short-range correlations are consistently included. The comparison of
results obtained in different kinematics confirms that resolution of different
final states in the $^{16}$O(${\vec e},e'{\vec p}p$)$^{14}$C reaction may act
as a filter to disentangle and separately investigate the reaction processes
due to short-range correlations and two-body currents and indicates that
measurements of the components of the outgoing  proton polarization may offer
good opportunities to study short-range correlations.
\end{abstract}
%\keywords{Keywords: Nuclear reactions, 
%$^{16}$O(${\vec e},e'{\vec p}p$)$^{14}$C, reaction mechanism, 
%two-nucleon knockout, NN correlations}
%%%%%%%%%%%%%%%%%%%%%%%%%%%%%

\section{Introduction}

It has always been a great challenge of nuclear physics to develop experiments 
and theoretical models able to explore short-range correlations (SRC). A
detailed investigation of these correlations, which are induced by the 
repulsive components of the nucleon-nucleon ($NN$) interaction, should give
insight into the structure of the interaction of two nucleons in the nuclear
medium.

Since a long time electromagnetically induced two-nucleon knockout reactions 
have been devised as a powerful tool for such investigation, since the 
probability that a real or virtual photon is absorbed by a pair of nucleons 
should be a direct measure of the correlations between these 
nucleons~\cite{Gottfried,Oxford}. In particular, the ($e,e'pp$) reaction has
been envisaged as the preferential tool for studying SRC in nuclei, since in
such a reaction the competing contributions of two-body currents are highly 
suppressed. Such triple coincidence experiments have been made possible only 
recently by the progress in accelerator and detector technology. First 
measurements of the exclusive $^{16}$O($e,e'pp$)$^{14}$C reaction have been 
performed at NIKHEF in Amsterdam~\cite{Gerco,NIKHEF,Ronald} and MAMI in 
Mainz~\cite{Rosner}. Investigations on these data
~\cite{Gerco,NIKHEF,Ronald,Rosner,sf} indicate that resolution of discrete 
final states provides an interesting tool to disentangle and thus separately 
investigate contributions of one-body currents, due to SRC, and two-body isobar
currents. In particular, direct and clear evidence for SRC has been obtained 
for the transition to the ground state of $^{14}$C. This result opens up good 
perspectives that further theoretical and experimental efforts on two-nucleon 
knockout reactions will be able to determine SRC. 

Good opportunities to increase the richness of information available from
two-nucleon knockout reactions is offered by polarization experiments. In fact,
only the use of polarization degrees of freedom allows one to obtain complete 
information on all possible reaction matrix elements. Reactions with polarized 
particles depend on a larger number of observables, which are hidden 
in the unpolarized case, where the sum and/or average over spin states is 
performed. These observables are represented by new response
functions~\cite{Oxford}, whose determination can impose more severe constraints
on theoretical models. In fact, some of these observables are expected to be 
sensitive to the small components of the transition amplitudes, which without
polarization are generally masked by the dominant ones. These small amplitudes 
often contain interesting information on subtle effects and may thus represent
a stringent test of theoretical models.  This is the place where polarization
observables enter, because in general they contain interference terms of the 
various matrix elements in different ways. Thus, a small amplitude may be 
considerably amplified by the interference with a dominant one. 

A complete experimental determination of the transition amplitudes requires a
long-term program involving the development of new experimental techniques and
many different polarization  measurements able to separate various sets of
structure functions. Unfortunately, some of these measurements are extremely
difficult in general and in particular in the case of two-nucleon knockout.
Therefore, a complete determination represents at present a too ambitious
programme.  

In order to exploit, at least partially, the potentiality of polarization
measurements, it is possible, however, to select and investigate specific 
situations corresponding to simpler experiments that appear feasible in the 
near future. They should allow one to gain a better control on the reaction
mechanism of two-nucleon knockout and hopefully study SRC. 

As a first step in the study of spin degrees of freeedom in electromagnetic
two-nucleon knockout reactions, we consider in this work nucleon recoil 
polarization. First theoretical predictions have been already presented in
Refs.~\cite{Gent1,Gent2} within the theoretical model developed by the Gent 
group. Here we give the most general formalism of (${\vec e},e'{\vec N}N$)
reactions in terms of structure functions and polarization observables.
Numerical results are shown for the specific case of the exclusive 
$^{16}$O(${\vec e},e'{\vec p}p$)$^{14}$C reaction for transitions to the 
lowest-lying states of the residual nucleus. Calculations have been performed
within the theoretical framework of Ref.~\cite{sf}. This model, that has been
successfully applied in the analysis of the first experimental cross
sections~\cite{NIKHEF,Ronald,Rosner}, gives a reasonably realistic base to our
numerical predictions. Results for the components of the outgoing proton
polarization and of the polarization transfer coefficient are presented and
discussed in different kinematics. Measurements of the proton recoil 
polarization require an efficient proton polarimeter and a double scattering. 
In addition, measurements of the polarization transfer coefficient require also
a polarized electron beam. Such experiments are therefore not easy, but seem 
reasonably within reach of available facilities. 

The general formalism is given Sec.~II and in Appendix. The numerical results 
are presented and discussed in Sec.~III. Some conclusions are drawn in Sec.~IV.

%%%%%%%%%%%%%%%%%%%%%%%%%%%%%%%%%

\section{Nucleon recoil polarization in 
(${\vec{\e}},\e'{\vec{\N}}\N$) reactions}

In general, the polarization of an outgoing nucleon, which is the 
expectation value of the spin, can be calculated as~\cite{Oxford}
\begin{equation} 
{\P} = \frac{ {\rm Tr} (M M^{\dag} {\ss})} {{\mathrm Tr} 
(M M^{\dag}) },
\label{eq:pol}
\end{equation}
where $M$ is the scattering amplitude of the reaction.

The coincidence cross section of an (${\vec e},e'{\vec N}N$) reaction, where 
two nucleons, with momenta $\p'_1$ and $\p'_2$ and energies $E'_1$ and $E'_2$, 
are emitted and the polarization of only one nucleon, with spin directed along 
$\hat{\s}$, is detected can be written
\begin{equation}
\frac{{\mathrm d}^5\sigma} {{\mathrm d}E'_0 {\mathrm d}\Omega'_0  \,
{\mathrm d}\Omega'_1 \/{\mathrm d}\Omega'_2 \/{\mathrm d}E'_1} = 
\sigma_0 \/ {\frac{1} {2}} \, \Big [1 + {\P}\/\cdot \hat {\s} 
+ h(A + {\P}\/'\cdot \hat{\s} )\Big ] ,
\label{eq:cs}
\end{equation}
where $E'_0$ is the energy of the outgoing electron, $\sigma_0$ the unpolarized 
differential cross section, ${\P}\/$ the outgoing nucleon polarization, $h$ the
electron helicity, $A$ the electron analyzing power and ${\P}\/'$ the 
polarization transfer coefficient.

The cross section and the polarization can be expressed in terms of the 
components of the hadron tensor
\begin{equation} 
W^{\mu \nu}_{\alpha \alpha '} = \overline{\sum_{\mathrm{i}}}
{\sum_{\mathrm{f}}} J^\mu _\alpha (\q) J^{\nu *} _{\alpha'} (\q) 
\delta (E_{\mathrm{i}} - E_{\mathrm{f}}) ,
\label{eq:htens}
\end{equation}
where $\alpha$ and $\alpha '$ are the eigenvalues of the spin of the considered
particle in the given reference frame and $\q$ is the momentum transfer. The
quantities $J^\mu_\alpha(\q)$ are the Fourier transforms of the transition
matrix elements of the nuclear charge-current density operator between initial
and final nuclear states
\begin{equation}
J^{\mu}({\mbox{\boldmath $q$}}) = \int \langle\Psi_{\rm{f}}
|\hat{J}^{\mu}({\mbox{\boldmath $r$}})|\Psi_{\rm{i}}\rangle
{\rm{e}}^{\,{\rm{i}}
{\footnotesize {\mbox{\boldmath $q$}}}
\cdot
{\footnotesize {\mbox{\boldmath $r$}}}
} {\rm d}{\mbox{\boldmath $r$}} .     \label{eq:jm}
\end{equation}

The symmetrical and antisymmetrical combinations of the components of the
hadron tensor produce nine spin dependent structure functions. The spin
dependence can be explicited in a spherical basis (see Appendix) as
\begin{equation} 
f_{\lambda\lambda'} = h^{\mathrm{u}}_{\lambda\lambda'} + {\hat{\s}} \cdot 
\h_{\lambda\lambda'} ,
\label{eq:sfunc}
\end{equation}
where ${\hat {\s}}$ is the unit vector in the spin space.  The structure
functions represent the response of the nucleus to the longitudinal and
transverse components of the electromagnetic interaction and only depend on the
energy and momentum transfer $\omega$ and $q$, the momenta of the two outgoing
nucleons $p'_1$ and $p'_2$ and the angles $\gamma_1$, between $\p'_1$ and $\q$,
$\gamma_2$, between $\p'_2$ and $\q$, and $\gamma_{12}$, between $\p'_1$ and
$\p'_2$. When the outgoing nucleon polarization is not detected and the cross
section is summed over the spin quantum numbers of the outgoing nucleon, the
spin independent structure functions $h^{\mathrm{u}}_{\lambda\lambda'}$ go over
to the structure functions $f_{\lambda\lambda'}$ of the unpolarized
case~\cite{Oxford,eep}. Thus, in this case only nine structure functions are
obtained. New structure functions are produced in the polarized case by the
components of $\h_{\lambda\lambda'}$ 

The components of the polarization and of the polarization transfer 
coefficient are obtained through the structure functions and, involving also 
the matrix elements of the lepton tensor $\rho_{\lambda\lambda'}$ and 
$\rho'_{\lambda\lambda'}$~\cite{Oxford}, are given by
\begin{equation} 
P^i = {\frac {\sum _{\lambda\lambda'} \rho  _{\lambda\lambda'} 
h^i _{\lambda\lambda'}} {\sum _{\lambda\lambda'} \rho  _{\lambda\lambda'}
 h^{\mathrm{u}}_{\lambda\lambda'}}} 
\label{eq:p1}
\end{equation}
and
\begin{equation} 
P'^i = {\frac {\sum _{\lambda\lambda'} \rho'  _{\lambda\lambda'} 
h'^i _{\lambda\lambda'}} {\sum _{\lambda\lambda'} 
\rho  _{\lambda\lambda'}h^{\mathrm{u}}_{\lambda\lambda'}}} .
\label{eq:p2}
\end{equation}

The vectors $\P$ and $\P'$ are usually projected onto the basis of unit 
vectors given by $\hat{\L}$\/ (parallel to the momentum ${\p}\/'$ of the 
outgoing particle), $\hat{\N}$\/ (in the direction 
of ${\q}\/\times{\p}\/'$) and $\hat{\S} = \hat{\N}\times\hat{\L}$\/, 
which define the cm helicity frame of the particle.

The explicit expressions of the cross section in terms of the structure
functions and of the structure functions in terms of the components of the
hadron tensor can be found in Appendix.

For the (${\vec e},e'{\vec N}N$) reaction in an unrestricted kinematics all the
components $P^N, P^L, P^S$ and $P'^N, P'^L, P'^S$ are allowed, as parity 
conservation does not impose in this case any restrictions. In this general
situation 36 structure functions are present: nine spin independent structure
functions in $\sigma_0$ ($h^{\mathrm{u}}$) and $A$ ($h'^{\mathrm{u}}$), and
nine spin dependent structure functions in each one of the components $i = N,
L, S$ of $\P$ ($h^i$) and $\P'$ ($h'^i$). 

This number is reduced in particular situations (see Appendix for more
details). When the angle $\alpha$ between the $\p'_1$ $\q$ plane and the 
electron scattering plane is equal to zero, all the 36 structure functions are in
general non-vanishing, but those of them which are multiplied by $\sin\alpha$ or
$\sin2\alpha$ in Eq.~(\ref{eq:cssec}) do not contribute to the cross section
and to the components of $\P$ and $\P'$. As a consequence, the condition
$\alpha = 0$ reduces to 24 the number of structure functions.

When the vectors $\q$, $\p'_1$, $\p'_2$ lie in the same plane, parity
conservation combined with the general properties of the hadron tensor reduces
the number of non-vanishing structure functions to 18: five in $\sigma_0$ and
$A$ and 13 in the components of $\P$ and $\P'$. This result is similar to
that obtained for the (${\vec e},e'{\vec N}$) reaction  in an unrestricted
kinematics~\cite{Oxford,eep}.                      

For a coplanar kinematics, i.e. when the initial and final electrons and the 
outgoing nucleons lie in the same plane, those of the 18 non-vanishing 
structure functions which are multiplied by $\sin\alpha$ or $\sin2\alpha$ do
not contribute and the number is reduced to 12, 4 spin independent and 8 spin
dependent, and only the components $P^N, P'^L$ and $P'^S$ survive. This result 
is similar to that obtained for the (${\vec e},e'{\vec N}$) reaction in a
coplanar kinematics~\cite{Oxford,eep}.                      

A particular situation occurs in the interesting case of the super-parallel 
kinematics, where the two outgoing nucleons are ejected parallel and
antiparallel to the momentum transfer. In this case only two structure 
functions, $h^{\mathrm{u}}_{00}$ and $h^{\mathrm{u}}_{11}$, do not vanish in 
the unpolarized cross section and three, $h^N_{01}, \overline{h}\/'^S_{01}, 
h'^L_{11}$, when polarization is considered. As a consequence, $P^N, P'^L$ and 
$P'^S$ are each directly proportional to only one structure function. This 
result is similar to the one obtained for the (${\vec e},e'{\vec N}$) reaction 
in parallel kinematics~\cite{Oxford,eep}.

It is worthwhile to investigate what happens when final-state interactions
(FSI) are neglected and the  plane-wave (PW) approximation is used for the
outgoing nucleons wave functions. 

The behaviour of the hadron tensor under time reversal and parity 
transformation has the property~\cite{Pick} 

\begin{equation}
W^{\mu\nu}({\s}, (-)) = W^{\nu\mu}(-{\s}, (+)) ,
\label{eq:pick}
\end{equation}
where ${\s}$ is the spin vector in the  ejectile rest
frame, and the dependence on the final state boundary condition for 
incoming $(-)$ and outgoing $(+)$ scattered waves is shown. For nucleon
knockout, the $(-)$ condition is appropriate. When the boundary conditions can 
be ignored, as in the PW approximation, Eq.~(\ref{eq:pick}) states that the 
symmetric part of $W^{\mu\nu}$ is independent of ${\s}$ and the antisymmetric 
part is proportional to ${\s}$. This is because, owing to the 
spin-${\frac{1}{2}}$ nature of the nucleon, the dependence of $W^{\mu\nu}$ on 
${\s}$ is at most linear. Therefore, in the PW approximation 
$P^N = P^L = P^S = 0$, while $\P'$ does not vanish.

%%%%%%%%%%%%%%%%%%%%%%%%%%%%%%%%%
\section{Proton recoil polarization in the 
$^{16}$O(${\vec{\e}},\e'{\vec{\p}}\p$)$^{14}$C reaction}

An experimental separation of the various structure functions would be of great
interest, but appears extremely difficult. A measurement of the nucleon recoil
polarization would be simpler and less affected by experimental errors, as it
is obtained through the determination of asymmetries. Therefore, in this 
section we present numerical predictions for the outgoing proton polarization 
$\P$ and the polarization transfer coefficient $\P'$ of the 
$^{16}$O(${\vec e},e'{\vec p}p$)$^{14}$C reaction leading to the lowest-lying 
discrete states in the residual nucleus. This reaction is of
particular interest for our investigation, due to the presence of discrete 
states in the experimental spectrum of the residual nucleus $^{14}$C, well
separated in energy and that can be separated with high-resolution experiments.
Cross sections calculations pointed out that transitions to different states 
can be differently affected by the two reaction processes due to SRC and 
two-body currents~\cite{sf}. Thus, the experimental separation of different
final states can act as a filter for the study of the two processes. Recent
experiments at NIKHEF and MAMI~\cite {Gerco,NIKHEF,Ronald,Rosner} have been 
able to resolve the lowest-lying states of $^{14}$C and have confirmed, in 
comparison with the theoretical results, the predicted selectivity of the 
exclusive reaction involving different transitions. In particular, clear 
evidence of the dominant contribution of SRC has been obtained for the 
transition to the 0$^+$ ground state, while the transition to the 1$^+$ state 
at 11.31 MeV appears better dominated by the $\Delta$ isobar current.   

New and complementary information is in principle available from polarization
observables. The aim of our investigation is to clarify the sensitivity of $\P$
and $\P'$ to the two competing reaction processes and in particular to SRC. 

Calculations have been performed within the same theoretical model~\cite{sf} 
used for the analysis of the available cross section data. A detailed 
description of the theoretical framework can be found in Ref.~\cite{sf} and in 
a series of previous papers where the different aspects of the model have been
developed~\cite{GP,eepp,Wouter,delta}. Here we summarize only the main 
features.  

The transition matrix elements $J^\mu(\q)$ in Eq.~(\ref{eq:jm}), for an 
exclusive reaction and under the assumption of a direct knockout mechanism, 
can be reduced to a form which contains three main ingredients: the two-nucleon 
overlap integral, the nuclear current and the final-state wave function of the
two outgoing nucleons.

In the calculations the scattering state is given by the product of two 
uncoupled single-particle distorted wave functions, eigenfunctions of a complex
phenomenological optical potential which contains a central, a Coulomb and a 
spin-orbit term. 

The nuclear current operator is the sum of a one-body and a two-body part.
These two parts correspond to the two reaction processes. In fact, while two
nucleons are naturally ejected by a two-body current, even if correlations are
not explicitly included in the two-nucleon wave function, they cannot be
ejected by a one-body current without correlations. Thus, the contribution of 
the one-body current is entirely due to correlations. In the calculations the 
one-body part contains a Coulomb, a convective and a spin term. For $pp$ 
knockout the two-body current, which is completely transverse, contains only
the contributions of non charge-exchange processes with intermediate
$\Delta$-isobar configurations in the intermediate state~\cite{sf,delta}. 

The two-nucleon overlap integrals are taken, for the different final states, 
from the calculation of the two-proton spectral function of 
$^{16}$O~\cite{Wouter}, where long-range and short-range correlations are 
consistently included. They are expressed in terms of a sum of products of 
relative and cm wave functions. Different components of relative and cm motion 
contribute to different transitions. They are weighed in the sum by two-proton 
removal amplitudes calculated within a large shell-model basis. SRC are 
included in the radial wave functions of relative motion through defect 
functions, defined by the difference between correlated and uncorrelated 
relative wave functions, and which are different for different relative states 
and for different $NN$ potentials. 

Therefore, SRC play a different role in different relative states. They are 
quite strong for the $^1S_0$  state and much weaker for $^3P_j$ states (the 
notation $^{2S+1}l_j$, for $l=S,P...$, is here used for the relative states). 
An opposite effect is given by the two-body $\Delta$ current, whose 
contribution is strongly reduced for $^1S_0$ $pp$ knockout, since there the 
generally dominant contribution of that current, due to the magnetic dipole 
$NN \leftrightarrow N\Delta$ transition, is suppressed~\cite{gnn,delta1}. Thus,
$^1S_0$ $pp$ knockout is generally dominated by SRC, while the $\Delta$ current
is more important in $^3P_j$ $pp$ knockout.

This result explains the above mentioned selectivity of the exclusive cross
sections involving the transitions to the 0$^+$ ground state and the 1$^+$
state. In fact, only $^3P$ relative states, $^3P_0$ $^3P_1$ $^3P_2$, contribute 
for the 1$^+$ state, whose cross section is generally dominated by the $\Delta$
current. The two relative waves $^1S_0$ and $^3P_1$ contribute for the 0$^+$ 
ground state and it is possible to envisage suitable kinematics where the role 
of $^1S_0$ $pp$ knockout and thus of SRC becomes dominant. 

In our investigation of polarization observables in the 
$^{16}$O(${\vec e},e'{\vec p}p$)$^{14}$C reaction calculations have been 
performed in different kinematics. In order to obtain a more complete 
information it is interesting to consider both coplanar and out-of-plane
kinematics. Experiments in coplanar kinematics are certainly simpler, but give
access only to the components $P^N$, $P'^L$ and 
$P'^S$, while all the components of ${\mbox{\boldmath $P$}}$ and 
${\mbox{\boldmath $P$}}'$ are present in an out-of-plane kinematics. 

A special and interesting case of coplanar kinematics is represented by the 
so-called super-parallel kinematics~\cite{GP}, where the knocked-out nucleons 
are detected parallel and anti-parallel to the transferred momentum $\q$. This
kinematics, that has been realized in the experiments at MAMI~\cite{Rosner},
is favored by the fact that only two structure functions, $f_{00}$
($h^{\mathrm{u}}_{00}$) and $f_{11}$ ($h^{\mathrm{u}}_{11}$), contribute to the
unpolarized cross section $\sigma_0$ and $P^N$, $P'^L$,
 $P'^S$ are each directly proportional to one structure function: 
$h^N_{01}$, $h'^L_{11}$, 
$\overline h'^S_{01}$, respectively.    

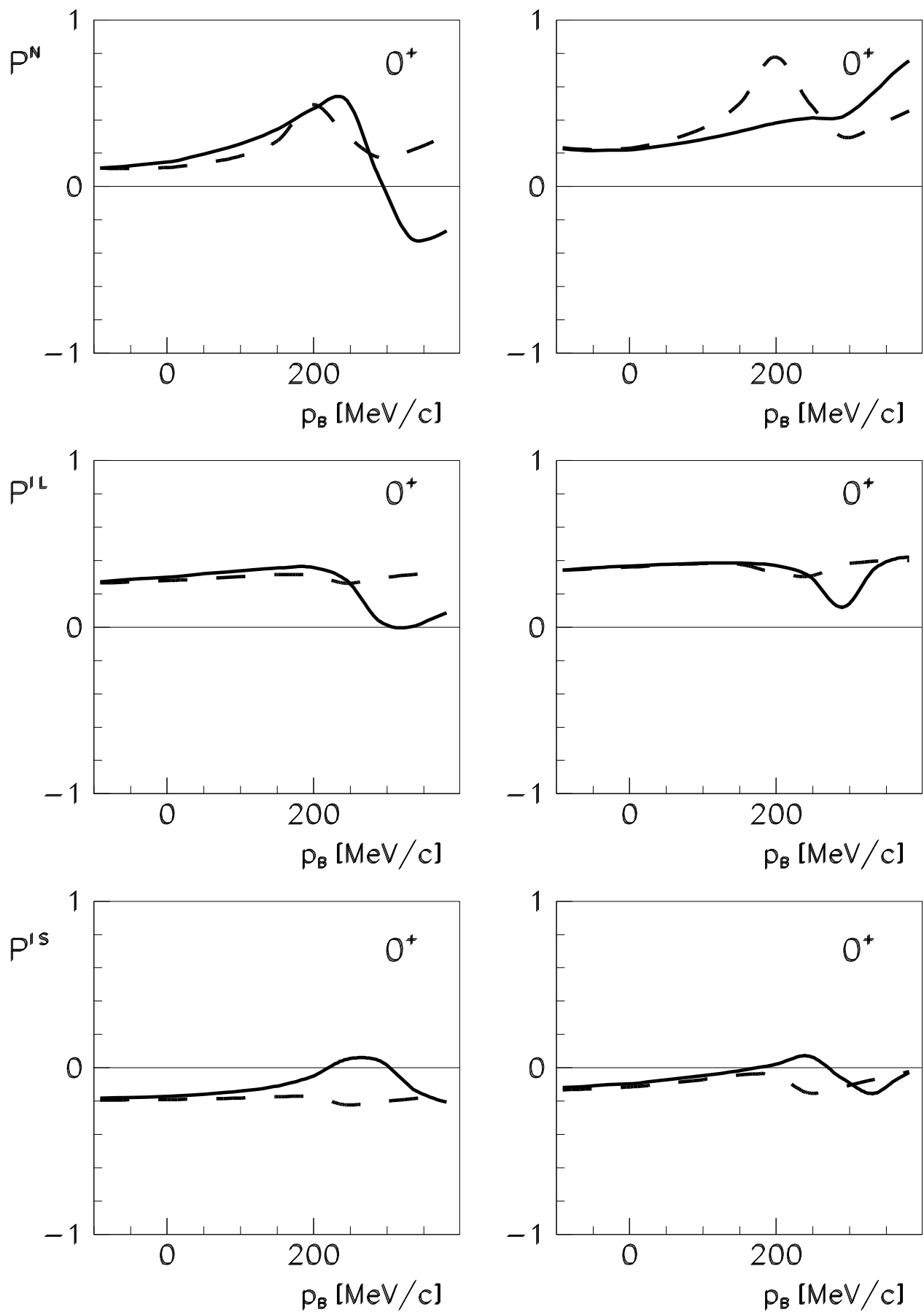
\begin{figure}
\caption[]{The components of the outgoing proton polarization $P^N$ 
and of the polarization transfer coefficient $P'^L$ and 
$P'^S$ of the 
$^{16}$O(${\vec e},e'{\vec p}p$)$^{14}$C reaction as a function of the recoil
momentum $p_{\mathrm{B}}$ for the transition $0^+$ ground state of $^{14}$C 
($E_{2\mathrm{m}} = 22.33$ Mev) in the super-parallel kinematics
($\gamma_1=0^{\mathrm{o}}$, $\gamma_2=180^{\mathrm{o}}$) with an incident
electron energy $E_{0} = 855$ MeV, $\omega = 215$ MeV and $q = 316$ MeV/$c$. 
Different values of  $p_{\mathrm{B}}$ are obtained changing the kinetic 
energies of the outgoing protons. Positive (negative) values of 
$p_{\mathrm{B}}$ refer to situations where 
${\mbox{\boldmath $p$}}_{\mathrm{B}}$ is parallel (antiparallel) to 
${\mbox{\boldmath $q$}}$. Polarization is considered for the proton
characterized by the momentum ${\mbox{\boldmath $p$}}'_1$. The dashed lines 
give the contribution of the one-body current, the solid line the sum of the 
one-body and the two-body $\Delta$-current. The defect functions for the 
Bonn-A and Reid $NN$ potentials are used in the left and right panels, 
respectively. The optical potential is taken from 
Ref.~\cite{Nad}.                 
\label{fig:fig1}
}
\end{figure}

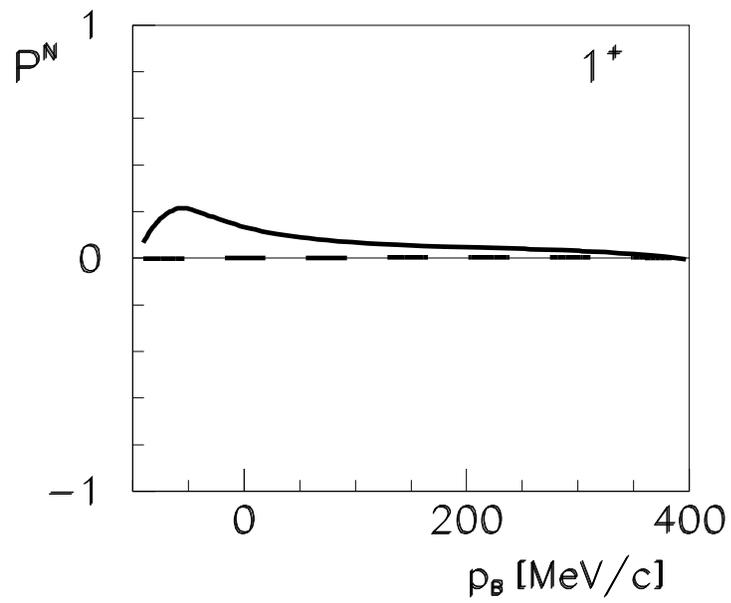
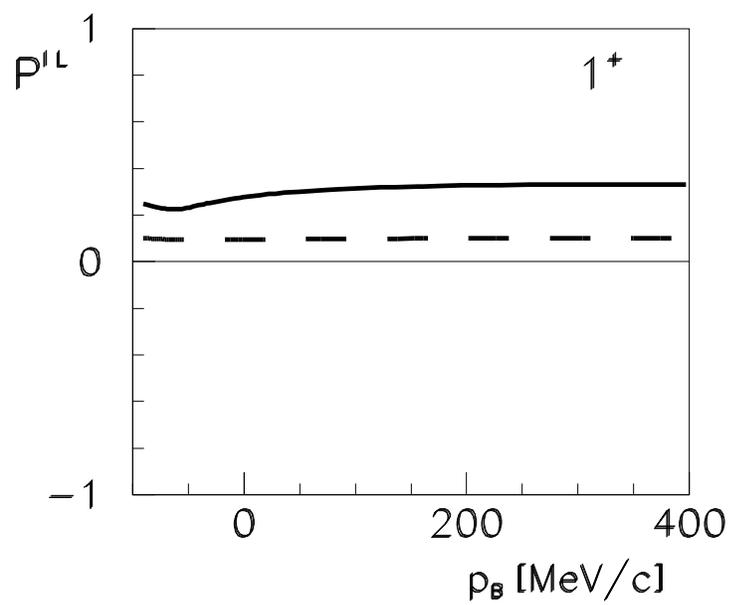
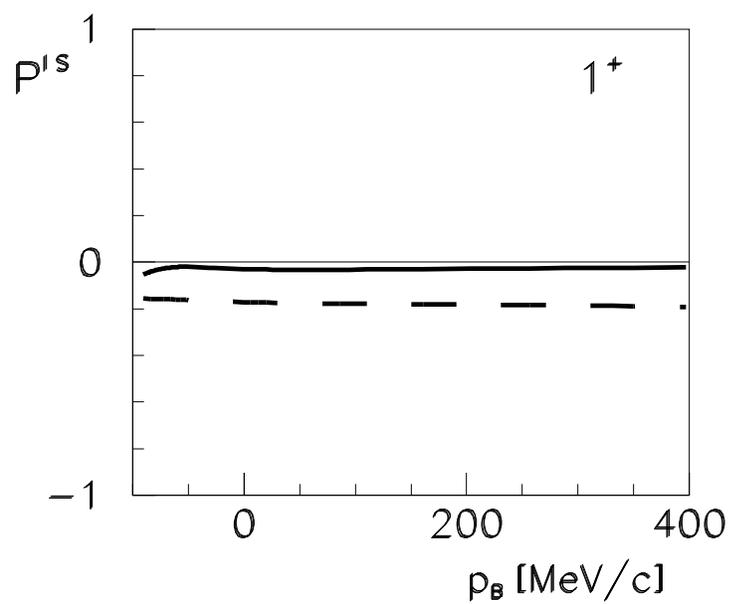
\begin{figure}
\caption[]{The components of the outgoing proton polarization $P^N$ 
and of the polarization transfer coefficient $P'^L$ and 
$P'^S$ of the 
$^{16}$O(${\vec e},e'{\vec p}p$)$^{14}$C reaction as a function of the recoil
momentum $p_{\mathrm{B}}$ for the transition $1^+$ state of $^{14}$C at 11.31 
MeV ($E_{2\mathrm{m}} = 33.64$ Mev) in the same kinematics and conditions and 
with same line convention as in Fig.~\ref{fig:fig1}. The defect functions for 
the Bonn-A $NN$ potential are used. 
\label{fig:fig2}
}
\end{figure}

The unpolarized differential cross section as well as the two unpolarized
structure functions in the super-parallel kinematics of the MAMI experiment
have been already shown and discussed in Ref.~\cite{sf}. The polarization
observables $P^N$, $P'^L$, $P'^S$ are 
displayed in Figs.~\ref{fig:fig1} and~\ref{fig:fig2} for the transitions to the
$0^+$ ground state and the $1^+$ state of $^{14}$C, respectively, as a function
of the recoil ($p_{\mathrm{B}}$) or  missing momentum ($p_{2{\mathrm{m}}}$), 
defined by 
\begin{equation}
\p_{2\mathrm{m}} = \p_{\mathrm{B}} =  \q -\p'_{1} - \p'_{2}.  
\label{eq:pm}
\end{equation}                         

In Figs.~\ref{fig:fig1} and ~\ref{fig:fig2} the result given by the sum of the
one-body and the two-body current is compared with the separate contribution
of the one-body current. In Fig.~\ref{fig:fig1}, for the transition to the 
ground state, where effects of SRC are expected to be more relevant, results 
given by the defect functions for the Bonn-A and Reid $NN$ potentials are 
compared.

In the calculations each state is characterized by a particular value of the
missing energy, given by
\begin{equation}
E_{2\mathrm{m}} = \omega - T'_{1} - T'_{2} -T_{\mathrm{B}} =
E_{\mathrm{s}} + E_{\mathrm{x}} , \label{eq:em}
\end{equation}
where $T'_{1}$, $T'_{2}$ and $T_{\mathrm{B}}$ are the kinetic energies of the
two outgoing nucleons and of the residual nucleus, respectively,
$E_{\mathrm{s}}$ is the separation energy at threshold for two-nucleon
emission and $E_{\mathrm{x}}$ is the excitation energy of the residual nucleus.
 
In the super-parallel kinematics all possible values of $p_{\mathrm{B}}$ are
explored, for a fixed value of the energy and momentum transfer and for a 
particular final state, changing the values of the kinetic energies of the 
outgoing nucleons.

An analysis of the results as a function of the recoil momentum appears of
particular interest. The shape of the unpolarized differential cross section 
is determined by the value of the cm orbital angular momentum $L$ of the 
knocked out pair~\cite{Gottfried,sf,eepp}. Different components of relative and
cm motion contribute to the two-nucleon overlap functions for each final state.
The shape of the calculated cross section is therefore driven by the component 
which gives the major contribution. For the $0^+$ state, the $^1S_0$ and 
$^3P_1$ relative waves are combined with $L= 0$ and $L = 1$, respectively; for 
the $1^+$ state the $^3P_0$, $^3P_1$ and $^3P_2$ relative waves are all 
combined with $L= 1$~\cite{sf}. Cross sections calculations in the considered 
kinematics indicate that the transition to the $0^+$ ground state has an $s$ 
wave shape, which is due to the major role played by $^1S_0$ $pp$ knockout, 
dominated by SRC, for low values of the recoil momentum. The $p$ wave 
component, which is due to the $^3P_1$ relative state and is dominated by the 
$\Delta$ current, becomes meaningful only at large values of $p_{\mathrm{B}}$, 
beyond $150-200$ MeV/$c$, where the contribution of the $s$ wave becomes much 
smaller~\cite{sf}. For the transition to the $1^+$ state the cross section has 
a $p$ wave shape and is almost entirely due to the $\Delta$ current~\cite{sf}. 
These theoretical findings have been confirmed in comparison with 
data~\cite{Rosner}. 

The polarization observables displayed in Fig.~\ref{fig:fig1} for the $0^+$ 
state confirm the dominant role of the one-body current up to about $150-200$ 
MeV/$c$. For larger values of $p_{\mathrm{B}}$, where the cross section is 
smaller, the contribution of the $\Delta$ current becomes more relevant. This 
is due to its interference with the one-body current, which is meaningful 
in the interference longitudinal-transverse structure functions 
$h^N_{01}$ and $\overline h'^S_{01}$ in this region of 
recoil momenta. For the component $P'^L$, which is proportional to 
the transverse polarized structure function $h'^L_{11}$, the 
separate contribution of the $\Delta$ current is large and of about the same 
size as that of the one-body current for all the considered values of 
$p_{\mathrm{B}}$. Thus, in this case, effects of interference between the 
one-body current and the two-body current are important over all the momentum 
distribution.  

The results given by the two sets of defect functions from Bonn-A and Reid 
potentials on the components of $P^N$, $P'^L$ and $P'^S$ are qualitatively 
similar. The numerical differences, however, are appreciable and even large at 
large values of $p_{\mathrm{B}}$, especially for the component $P^N$. 
This effect is predominantly due to the different interference between the 
one-body current and the $\Delta$ current. 

The results in Fig.~\ref{fig:fig1} for the transitions to the $0^+$ state 
indicate that in the super-parallel kinematics the considered polarization 
observables are sizable and sensitive to effects of SRC. The component of the
proton recoil polarization $P^N$ appears very well suited to study correlation 
effects. A  measurement of this component would be simpler than a measurement 
of $P'^L$ and $P'^S$, which would require also a polarized electron beam. On 
the other hand, since all the components of $\P$ vanish in the PW 
approximation, $P^N$ could be also sensitive to the treatment of FSI. 

The polarization observables displayed in Fig.~\ref{fig:fig2} for the 
transition to the $1^+$ state are smaller and, as expected, more sensitive to 
the contribution of the $\Delta$ current. The results in Figs.~\ref{fig:fig1} 
and~\ref{fig:fig2} confirm that different final states in the exclusive 
$^{16}$O($e,e'pp$)$^{14}$C reaction may act as a filter for the study of the
two reaction processes due to SRC and two-body currents. 

The super-parallel kinematics represents a special case of coplanar kinematics
where only 5 structure functions contribute to the unpolarized and polarized
cross sections. In general, in a coplanar kinematics 12 structure functions are
present (see Sec.~II and Appendix): 4 are contained in the unpolarized cross
section $\sigma_0$ and 8 in the surviving components of the of the proton 
recoil polarization $P^N$ and of the polarization transfer coefficient $P'^L$ 
and $P'^S$.

As  an example, we have here considered a specific kinematical setting included 
in the first experiment on $^{16}$O carried out at NIKHEF~\cite{Gerco}, with an 
incident electron energy of 584 MeV, $\omega = 212$ MeV and $q = 300$ MeV/$c$. 
The kinetic energy of the first outgoing proton $T'_1$ is 137 MeV and the angle
$\gamma_1 = 30^{\mathrm{o}}$, on the opposite side of the outgoing electron 
with respect to the momentum transfer. Changing the angle $\gamma_2$ on the 
other side, different values of the recoil momentum $\p_{\mathrm{B}}$ are 
explored in the range between $-250$ and $300$ MeV/$c$, including the zero 
values at $\gamma_2 \simeq 120^{\mathrm{o}}$~\cite{sf}. 

The unpolarized cross section has been already discussed in Ref.~\cite{sf}
and is shown again, for the transition to the $0^+$ ground state of $^{14}$C, in
the top panels of Fig.~4. The qualitative features are similar to those 
obtained in the super-parallel kinematics of the MAMI experiment. For the
$0^+$ state the shape of the recoil momentum or angular distribution is driven
by the component with $L=0$ , that is by $^1S_0$ $pp$ knockout, and is thus
dominated by SRC at low values of $p_{\mathrm{B}}$. The contribution with $L =
1$, combined with the $^3P_1$ relative state, which is better driven by the 
$\Delta$ current, is negligible when $p_{\mathrm{B}}$ is low, but becomes 
meaningful at large values of the recoil momentum, where the contribution of the
$s$ wave component is strongly suppressed. In contrast, the transition to the 
$1^+$ state, which contains only $^3P$ relative waves and $L = 1$ cm components,
has a $p$ wave shape and is dominated by the $\Delta$ current. 

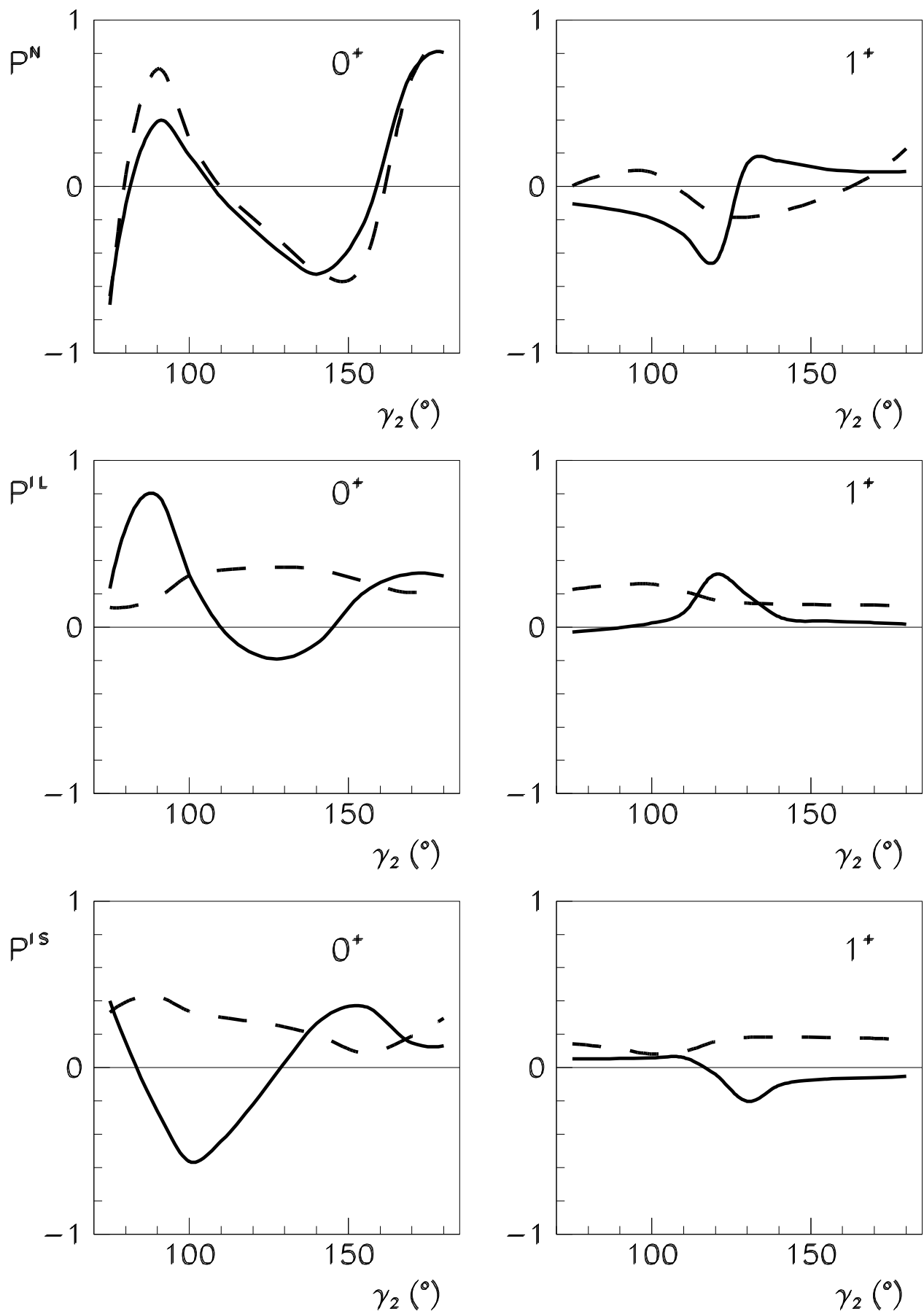
\begin{figure}
\caption[]{The components of the outgoing proton polarization $P^N$ 
and of the polarization transfer coefficient $P'^L$ and 
$P'^S$ of the 
$^{16}$O(${\vec e},e'{\vec p}p$)$^{14}$C reaction as a function of the angle 
$\gamma_2$ for the transitions to the $0^+$ ground state (left panels) and the 
$1^+$ state at 11.31 MeV (right panels) of $^{14}$C. The incident electron 
energy is $E_{0} = 584$ MeV, $\omega = 212$ MeV and $q = 300$ MeV/$c$, 
$T'_1 = 137$ MeV and $\gamma_1 = 30^{\mathrm{o}}$, on the opposite side of the
outgoing electron with respect to the momentum transfer. Polarization is 
considered for the proton characterized by the kinetic energy $T'_1$ and angle 
$\gamma_1$. The defect functions for the Bonn-A potential and the optical
potential of Ref.~\cite{Nad} are used. 
\label{fig:fig3}
}
\end{figure}

The polarization observables $P^N$, $P'^L$, and $P'^S$ are displayed, as a 
function of the angle $\gamma_2$, in Fig.~\ref{fig:fig3} for the transitions to
the $0^+$ ground state and to the $1^+$ state of $^{14}$C. For the $0^+$ state 
$P^N$ is large and dominated over all the angular distribution by the one-body 
current and thus by SRC. The components of the polarization transfer 
coefficient $P'^L$ and $P'^S$ are also large, but appear better driven by the 
$\Delta$ current. Thus, also in this kinematics the component $P^N$ turns out 
to be very well suited to study SRC, while in $P'^L$ and $P'^S$ the 
contribution of the $\Delta$ current is relevant and intertwined with that of 
the one-body current. Therefore, for these two components, whose measurement 
requires more complicated experiments, the separation of either contribution of
the two reaction processes appears difficult.

The comparison between the results with  the two kinematical settings in 
Figs.~\ref{fig:fig1} and \ref{fig:fig3} for the $0^+$ state indicates that in 
Fig.~\ref{fig:fig3}, where a larger number of structure functions contribute, 
the polarization observables are generally larger and the role of SRC in $P^N$ 
is dominant over all the distribution. From this point of view, this kinematics
seems better suited for the study of SRC with polarization measurements. On the
other hand, the role of the $\Delta$ current in $P'^L$ and $P'^S$ is more 
relevant than in the super-parallel kinematics of Fig.~\ref{fig:fig1}.

Also in the kinematics of Fig.~\ref{fig:fig3} the polarization observables for 
the transition to the $1^+$ state turn out to be generally smaller than for 
the $0^+$ state. For this transition $P^N$, $P'^L$ and $P'^S$ are sizable only 
in the region of low vales of $p_{\mathrm{B}}$, where the cross section has the
minimum, and are driven over all the distribution by the $\Delta$ current. This
result confirms that the transition to the $1^+$ state is dominated by two-body
currents and indicates that  polarization measurements for this final state 
appear more difficult. 

The results of Fig.~\ref{fig:fig3} have been obtained with the defect functions
from the Bonn-A potential. Calculations performed with the set of defect 
functions from the Reid potential give appreciable differences, especially for 
the $0^+$ state and for the observable $P^N$, which is most affected by 
correlations. The main qualitative features of the results remain however 
unchanged with respect to those presented in Fig.~\ref{fig:fig3}.

All the components $N, L, S$ of $\P$ and $\P'$ can be explored with an
out-of-plane kinematics. Such kinematics can be realized when the angle
$\alpha$, between the $\p'_1$ $\q$ plane and the electron scattering plane, or
the azimuthal angle $\phi$ of the outogoing proton whose polarization is not
considered, or even both $\alpha$ and $\phi$  are different from zero. In the 
most general case, where $\alpha \neq 0$ and $\phi \neq 0$, 36 structure 
functions are obtained (see Sec.~II and Appendix). When $\phi \neq 0$, all these
structure functions are in general non-vanishing, but if $\alpha = 0$ those of 
them which are multiplied by $\sin\alpha$ or $\sin2\alpha$ do not give any 
contributions. Thus only 24 structure functions are active. If $\phi = 0$, the
vectors $\q$, $\p'_1$ and $\p'_2$ lie in the same plane and the structure
functions are the same as in coplanar kinematics, but if $\alpha \neq 0$ also 
those of them which are multiplied by $\sin\alpha$ or $\sin2\alpha$ contribute.
This gives 18 structure functions.

In order to explore all the components of $\P$ and $\P'$, two examples of
out-of-plane kinematics are here considered: $\alpha = 45^{\mathrm{o}}$ 
$\phi = 0^{\mathrm{o}}$ and $\alpha = 0^{\mathrm{o}}$ $\phi = 30^{\mathrm{o}}$.
The other kinematical variables are taken as in the coplanar kinematics of 
Fig.~\ref{fig:fig3}. Calculations have been performed only for the transition 
to the $0^+$ ground state, where effects of SRC are known to be most relevant.  

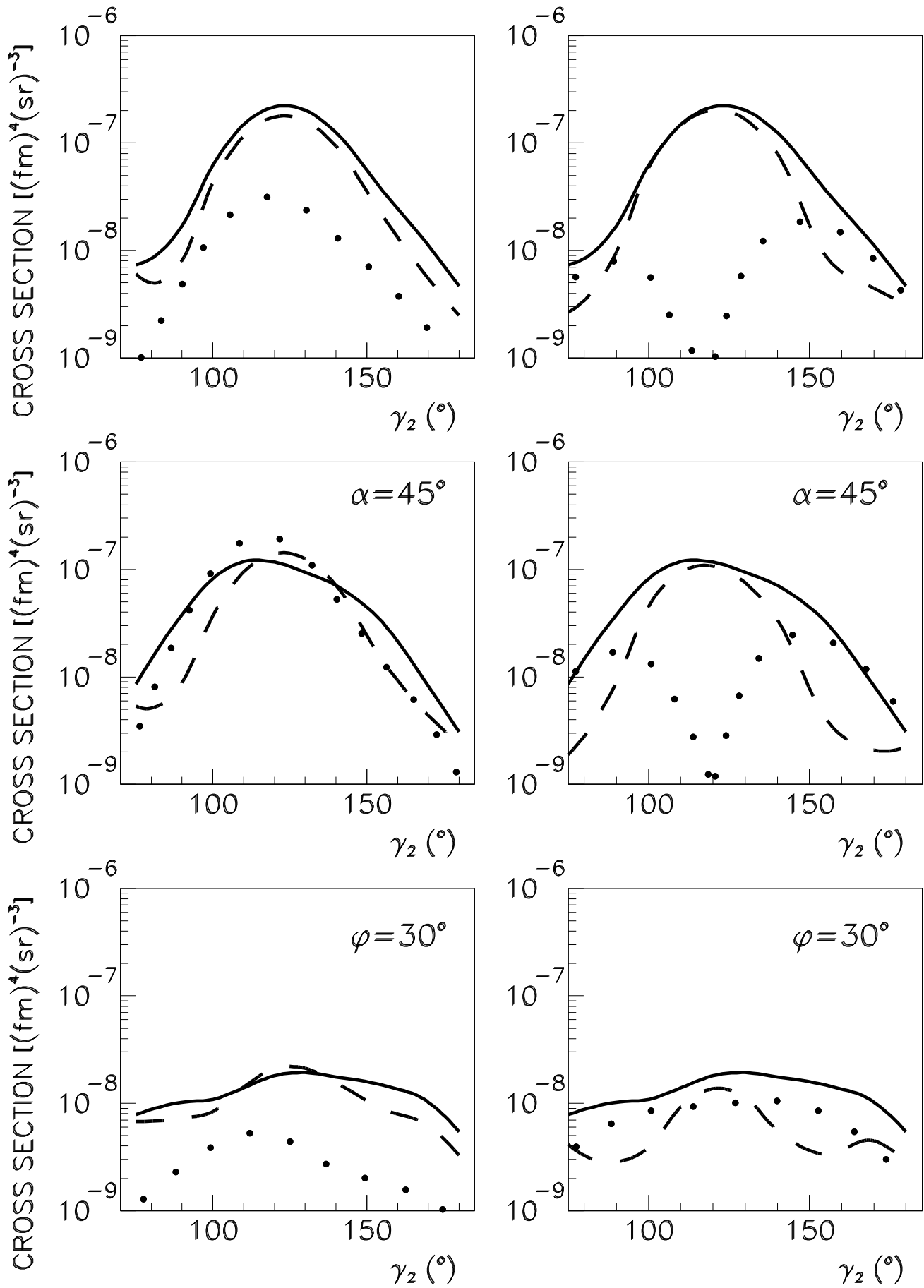
\begin{figure}
\caption[]{The differential cross section  of the $^{16}$O($e,e'pp$)$^{14}$C
reaction as a function of the angle $\gamma_2$ for the transition to the $0^+$
ground state of $^{14}$C in the same coplanar kinematics as in 
Fig.~\ref{fig:fig3} (top panels) and in two different out-of-plane kinematics 
with $\alpha =45^{\mathrm{o}}$ and with the azimuthal angle of the second  
outgoing proton $\phi =30^{\mathrm{o}}$. Defect functions and optical
potential as in Fig.~\ref{fig:fig3}. The separate contributions of the 
one-body and the two-body $\Delta$ currents are shown by the dashed and dotted
lines, respectively in the left panels. Separate contributions of the different
partial waves of relative motion $^1S_0$ and $^3P_1$ are shown by the dashed
and dotted lines, respectively, in the right panels. The solid lines are the
same in the left and right panels and give the final result sum of the various
contributions. 
\label{fig:fig4}
}
\end{figure}

The differential cross sections in the two considered out-of-plane kinematics 
are displayed in Fig.~\ref{fig:fig4} as a function of the angle $\gamma_2$ and
compared with the cross section of the corresponding coplanar kinematics. The
final results are compared in the three cases with the separate contributions
of the one-body and the two-body currents and of the $^1S_0$ and $^3P_1$ 
components of relative motion.  

In the kinematics with $\alpha = 45^{\mathrm{o}}$ the size of the cross
section is only slightly lower than in the coplanar kinematics. Also the shapes
of the curves given by the separate contributions are similar to those obtained
in the coplanar situation. The contribution of the $^1S_0$ component is,
however, reduced and that of $^3P_1$ enhanced. As a consequence, the role of 
$^3P_1$ $pp$ knockout, dominated by the $\Delta$ current, is enhanced in the
final cross section. This effect slightly changes the shape of the final
distribution. On the other hand, when $\alpha = 45^{\mathrm{o}}$ the role of
the $\Delta$ current becomes much more important also on the $^1S_0$ component,
as can be seen from the comparison between the results in the left and right
panels. Therefore, in the final cross section the contribution of the $\Delta$
current becomes competitive with that of the one-body current even for the
transition to the $0^+$ state. This is the main difference with respect to the
result of the coplanar kinematics and is due to the different role played by
the structure functions which are multiplied by factors including the angle 
$\alpha$. In particular, a significant role is played by the structure functions
multiplied by $\sin\alpha$ and $\sin2\alpha$, which do not contribute in the
coplanar kinematics, and where the effect of the $\Delta$ current is important.

In the kinematics with $\phi = 30^{\mathrm{o}}$ the cross section is about an
order of magnitude lower than in the peak region of the coplanar kinematics and
has a different shape. In practice, the peak has disappeared and there is no
more evidence of an $s$ shape distribution. In fact, the two contributions of 
$^1S_0$ and $^3P_1$ turn out to be of comparable size in this case. In spite of
that, the major role is still played by the one-body current, whose
contribution is relevant, even though not dominant, also on the $^3P_1$ relative
wave. A part of the differences with respect to the results of the coplanar
kinematics is due to the different structure functions and to their dependence
on the angle $\phi$. The main reason of the differences, however, can be
attributed to a kinematic effect. In fact, when the outgoing nucleon is taken
out of the plane, different values of the recoil momentum $p_{\mathrm{B}}$ are
obtained in the considered range of $\gamma_2$. In particular, low values in
the range between $-160$ and $160$ MeV/$c$ are forbidden. Thus, the main source
of difference with respect to the coplanar situation is that the region of
momenta between $-160$ and $160$ MeV/$c$, where the $s$ wave has the maximum
and the $p$ wave a minimum, has been cut. This explains the different shape of
the angular distributions in the top and bottom panels of Fig.~\ref{fig:fig4}. 

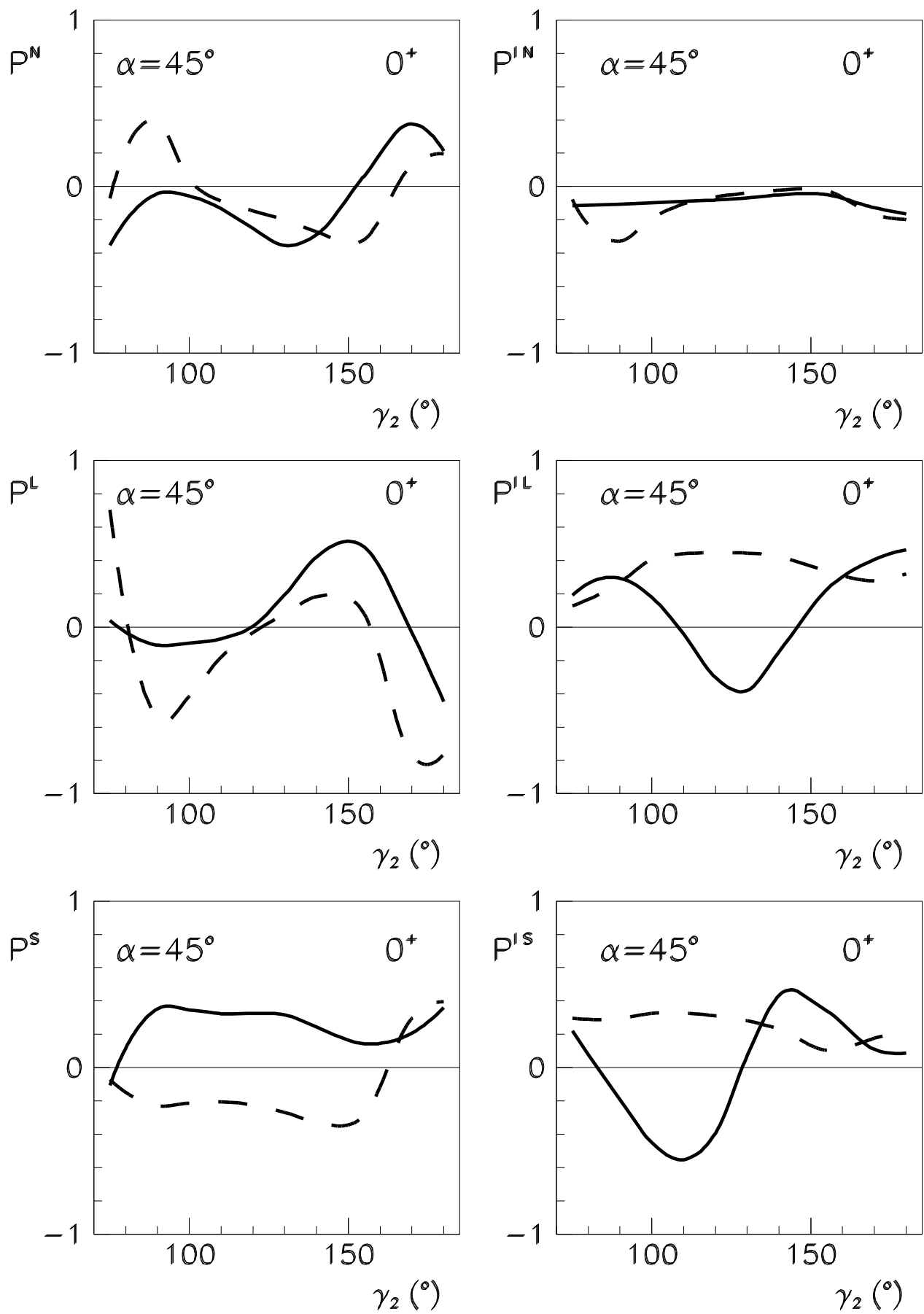
\begin{figure}
\caption[]{The components of the outgoing proton polarization 
${\mbox{\boldmath $P$}}$ (left panels) and of the polarization transfer 
coefficient ${\mbox{\boldmath $P$}}'$ (right panels) for the same reaction and
in the same out-of-plane kinematics with $\alpha =45^{\mathrm{o}}$ and with the
same conditions as in Fig.~\ref{fig:fig4}. Line convention as in 
Fig.~\ref{fig:fig1}. Polarization is considered for the proton characterized by
the momentum ${\mbox{\boldmath $p$}}'_1$. 
\label{fig:fig5}
}
\end{figure}

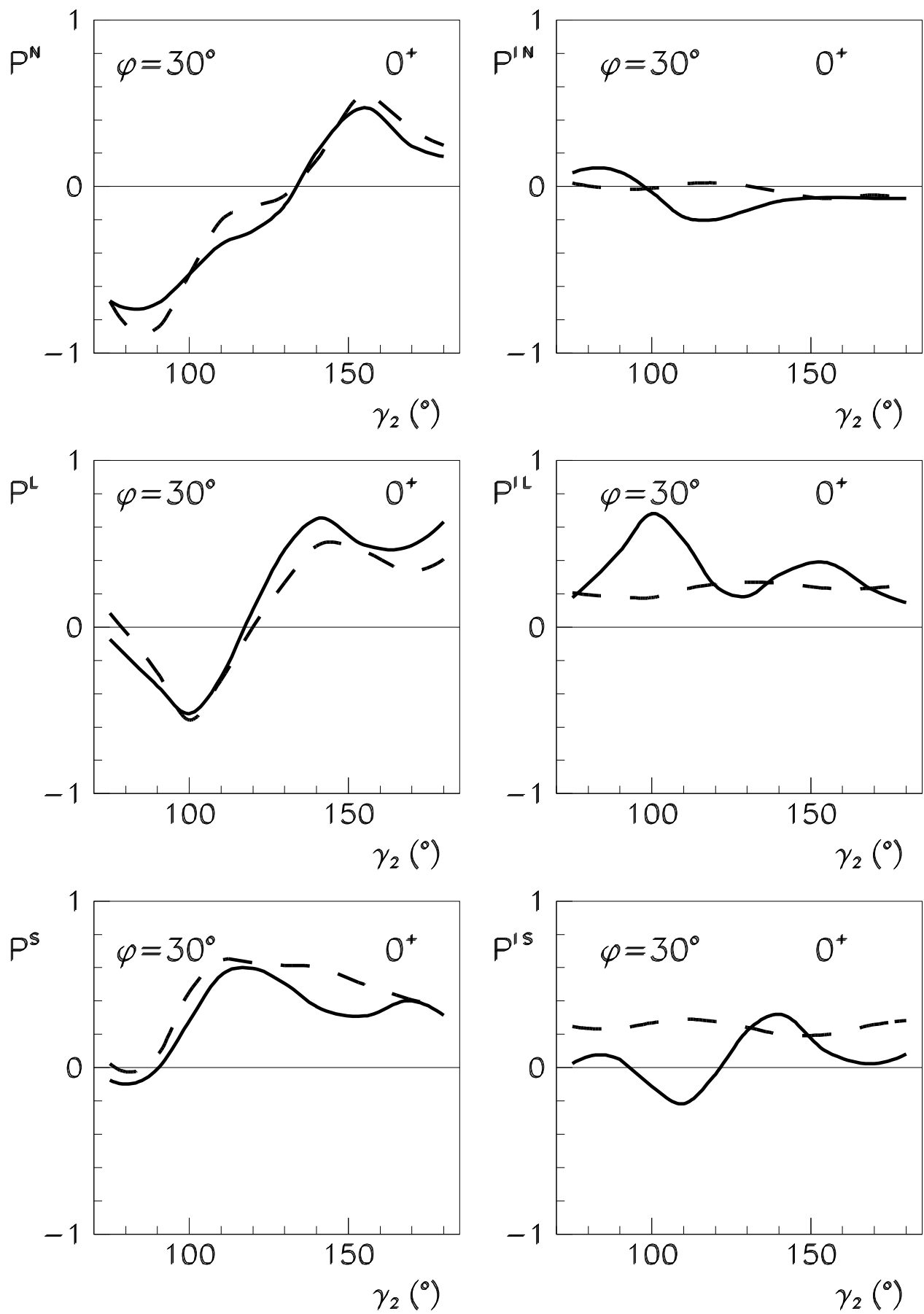
\begin{figure}
\caption[]{The components of the outgoing proton polarization 
${\mbox{\boldmath $P$}}$ (left panels) and of the polarization transfer 
coefficient ${\mbox{\boldmath $P$}}'$ (right panels) for the same reaction and
in the same out-of-plane kinematics with $\phi =30^{\mathrm{o}}$ and with the
same conditions as in Fig.~\ref{fig:fig4}. Line convention as in 
Fig.~\ref{fig:fig1}. Polarization is considered for the proton
characterized by the momentum ${\mbox{\boldmath $p$}}'_1$. 
\label{fig:fig6}
}
\end{figure}

The components of $\P$ and $\P'$ in the two kinematics with $\alpha =
45^{\mathrm{o}}$ and $\phi = 30^{\mathrm{o}}$ are displayed in 
Figs.~\ref{fig:fig5} and ~\ref{fig:fig6}, respectively.

In Fig.~\ref{fig:fig5}, with $\alpha = 45^{\mathrm{o}}$, the components $P^N$,
$P'^L$ and $P'^S$, already present in coplanar kinematics, are significantly
different from those displayed in Fig.~\ref{fig:fig3} for the same transition.
The differences are larger for $P^N$ and $P'^L$. The $\Delta$ current gives the
major contribution to $P'^L$ and $P'^S$. The one-body current gives the main
contribution to $P^N$, but in this kinematics the effect of the $\Delta$
current is meaningful also on this polarization component and larger than in
coplanar kinematics. Of the three observables $P'^N$, $P^L$ and $P^S$, which are
present only in an out-of-plane kinematics, $P'^N$ is small, while $P^L$ and 
$P^S$ are sizable. However, both contributions of SRC and two-body currents are
large and intertwined in these polarization components. Therefore, the
kinematics with $\alpha = 45^{\mathrm{o}}$  does not seem particularly well
suited to disentangle the two reaction processes and study SRC.

Correlation effects are much more important in  Fig.~\ref{fig:fig6}, for the
kinematics with $\phi = 30^{\mathrm{o}}$. In this case all the components of
the outgoing proton polarization $\P$ are sizable and driven by the one-body
current, which gives the main contribution also to $P'^N$. This component of
the polarization transfer coefficient, however, is small also in this
kinematics. The other components $P'^L$ and $P'^S$ are large, but also in
this case strongly affected by both contributions of the two reaction processes
due to SRC and two-body currents. 
       
%%%%%%%%%%%%%%%%%%%%%%%%%%%%%%%%%%%%%%%%
\section{Summary and conclusions}

In this paper we have discussed the general properties of the nucleon recoil
polarization in the (${\vec e},e'{\vec N}N$) reaction. In the most general 
situation no restrictions are due to parity conservation and 36 structure 
functions are active, more than in the (${\vec e},e'{\vec N}$) reaction, where
parity conservation reduces the number of structure functions available in an
unconstrained kinematics to 18. The same formal situation is obtained in 
(${\vec e},e'{\vec N}N$) only when the momenta of the two outgoing nucleons and
the momentum transfer lie in the same plane. A minor number of structure
functions can be obtained in more restricted kinematics: 12 in a coplanar
kinematics, as in the coplanar kinematics of (${\vec e},e'{\vec N}$), and 5 in
the special case of the super-parallel kinematics, as in the parallel 
kinematics of (${\vec e},e'{\vec N}$).

An experimental determination of the structure functions would be of great
interest, but appears at present extremely difficult. In order to exploit the 
potentiality of polarization measurements to increase the richness of 
information available in two-nucleon knockout reactions, it is anyhow possible
to envisage other interesting polarization experiments which appear more 
feasible. As an example, a measurement of the nucleon recoil polarization,
which is obtained through the determination of asymmetries, should be 
reasonably within reach of available facilities. 

A complete investigation of all the components of the outgoing nucleon
polarization $\P$ and of the polarization transfer coefficient $\P'$, which
implies also a polarized electron beam, requires out-of-plane experiments. In
the simpler case of a coplanar kinematics only the components $P^N$, $P'^L$ and
$P'^S$ are available, as in the coplanar kinematics of the 
(${\vec e},e'{\vec N}$) reaction. Moreover, as in (${\vec e},e'{\vec N}$), all
the components of the outgoing nucleon polarization $\P$ vanish in the PW
approximation.

The aim of this work was to explore the capability of polarization
measurements to disentangle and separately investigate the two reaction
processes due to correlations and two-body currents. With this aim and within
our theoretical framework, we have checked the sensitivity of $\P$ and $\P'$ to
the two competing processes. 

Numerical predictions have been presented for the specific case of the
exclusive $^{16}$O(${\vec e},e'{\vec p}p$)$^{14}$C reaction leading to the
lowest-lying discrete states of the residual nucleus. Our results confirm that 
the contribution of SRC as compared with that of the $\Delta$ current depends 
on the final state of the residual nucleus and that the transition to the 
ground state of $^{14}$C is particularly sensitive to correlation effects. 
Calculations performed in the coplanar kinematical settings realized for the 
cross section measurements at NIKHEF and MAMI indicate that for the transition 
to the ground state the major contribution to the only surviving component of 
the outgoing proton polarization, perpendicular to the plane, is given by 
correlations, while effects of two-body currents are almost negligible. In 
contrast, these effects are much more important or even dominant on the 
components of the polarization transfer coefficient.

In more complicated non coplanar kinematics the number of available observables
increases, but their sensitivity to nuclear correlations is reduced. It is 
possible to envisage particular conditions where correlation effects are 
dominant, but they correspond to situations where measurements are more 
difficult and cross sections generally smaller. 

In conclusion, we believe that a combined measurement of cross sections and
polarizations in (${\vec e},e'{\vec p}p$) reactions would provide a unique tool
to determine short-range nuclear correlations and could positively contribute
to clarify the behaviour of the short-range interaction of nucleons in the
nuclear medium. 

The analysis of polarization observables can be extended to other 
electromagnetic two-nucleon knockout reactions, such as ($e,e'pn$), 
($\gamma,pn$) and ($\gamma,pp$). 

Recent calculations~\cite{pn} have shown that the cross section of the 
exclusive $^{16}$O(${\vec e},e'{\vec p}n$)$^{14}$N reaction are sensitive to 
details of the nuclear correlations considered and in particular to the 
presence of the tensor component. Therefore, a study of polarization 
observables in the ($e,e'pn$) reaction would be of particular interest for the 
study of tensor correlations. 

Cross sections and photon asymmetries calculated for both ({$\gamma,pp$}) and
($\gamma,pn$) knockout reactions are dominated by two-body currents and only
slightly affected by correlation effects~\cite{gnn}. Thus, reactions induced by
real photons do not seem particularly well suited to study correlations, but
are anyhow interesting to give complementary information on other theoretical 
ingredients, for instance on the nuclear currents. Moreover, polarization 
observables could amplify the role of small amplitudes and of subtle effects 
hidden in the unpolarized case. Therefore, also ($\gamma,pp$) and ($\gamma,pn$)
reactions deserve a careful investigation.

%%%%%%%%%%%%%%%%%%%%%%%%%%%%%%%%%%%%%%%%
\section*{Appendix}

The components of the hadron tensor $W^{\mu\nu}$ are to be restricted by the
conditions of current conservation
\begin{equation} 
W^{\mu \nu} q_\mu = W^{\mu \nu} q_\nu =  0.
\label{eq:cc}
\end{equation}
Therefore, the sixteen components of the tensor are reduced to nine independent
quantities. By separating the symmetrical and antisymmetrical terms, one
obtains six symmetrical and three antysimmetrical independent components.

A spherical basis can be used and defined by the four-vectors 
\begin{equation} 
\epsilon^\mu_{\pm1} = \mp {\frac{1}{\sqrt{2}}} \, (0,1,\pm{\mathrm{i}},0), 
\,\,\,\, \, 
\epsilon^\mu_0 = \left ( {\frac{|\q|}{Q}},0,0,{\frac{\omega}{Q}} \right ),
\label{eq:sph}
\end{equation}
where $Q^2 = |\q|^2 -\omega^2$. (Note that this spherical basis is consistent
with Ref.~\cite{Oxford}, and is different from the one used in other papers,
where $\epsilon^\mu_0 = (1,0,0,0)$.)

Then, nine structure functions are obtained as a function of the hadron tensor
components~\cite{Oxford,GP}. Their expressions are given in Table~1, in the
reference frame where the $\z$ axis is taken parallel to $\q$ and the momentum
$\p'_1$ lies in the $\x \z$ plane.

The triple coincidence cross section for the electron induced reaction where two
nucleons are emitted is obtained from the contraction between the lepton tensor
and the hadron tensor as a linear combination of the nine structure
functions~\cite{Oxford,GP}
\begin{eqnarray} 
\frac{{\mathrm d}^5\sigma} {{\mathrm d}E'_0 {\mathrm d}\Omega'_0  \,
{\mathrm d}\Omega'_1 \/{\mathrm d}\Omega'_2 \/{\mathrm d}E'_1} = 
 \sigma_{\mathrm{M}} & \Omega_{\mathrm{f}} & f_{\mathrm{rec}}\, 
\sum_{\lambda\lambda'} \{ \rho_{\lambda\lambda'} f_{\lambda\lambda'} + h 
\rho'_{\lambda\lambda'} f'_{\lambda\lambda'} \} \nonumber \\
=  K & \Omega_{\mathrm{f}}&  f_{\mathrm{rec}} \Big \{ \epsilon_{\mathrm{L}} 
f_{00} + f_{11} + \sqrt{\epsilon_{\mathrm{L}}(1+\epsilon)} (f_{01} \cos\alpha + 
\overline f_{01} \sin\alpha) \nonumber \\
 - & \epsilon & (f_{1-1} \cos2\alpha + \overline f_{1-1} 
\sin2\alpha) \nonumber \\
  + & h & \Big [ {\sqrt{\epsilon_{\mathrm{L}} (1-\epsilon)}} 
(f'_{01} \sin\alpha + \overline f'_{01} \cos\alpha) + \sqrt{1-\epsilon^2} 
f'_{11} \Big ] \Big \}, \label{eq:cssec}
\end{eqnarray} 
where $\sigma_{\mathrm{M}}$ is the Mott scattering cross section,
\begin{equation} 
\Omega _{\mathrm{f}} = |\p'_1| E'_1 |\p'_2| E'_2,
\label{eq:of}
\end{equation}
\begin{equation} 
f_{\mathrm{rec}}^{-1} = 1 - \frac{E'_2}{E_{\mathrm{r}}} \frac{\p'_2 \cdot
\p_{\mathrm{r}}}{|\p'_2|^2},
\label{eq:rec}
\end{equation}
where $E_{\mathrm{r}}$ and $\p_{\mathrm{r}}$ are the relativistic energy and
momentum of the residual nucleus,
\begin{equation}
K = \frac{e^4}{16\pi^2} \, \frac{E'_0}{Q^2 E_0 (1-\epsilon)},
\label{eq:k}
\end{equation}
\begin{equation}
\epsilon = \left( 1 + 2 \frac {\vert \q \vert^2} {Q^2} \tan^2 \frac {\theta}
{2} \right)^{-1} ,
\label{eq:eps}
\end{equation}
\begin{equation}
\epsilon_{\mathrm L} = \frac {Q^2} {\vert \q \vert^2} \epsilon
\label{eq:epsL}
\end{equation}
and $E_0$ and $E'_0$ are the energies of the incident and outgoing electrons.
The components of the lepton tensor $\rho_{\lambda \lambda'}$ and
 $\rho'_{\lambda \lambda'}$ can be deduced from Eq.~(\ref{eq:cssec}).

The structure functions depend on the kinematical variables of the 
particular process under investigation. Even if a large number of variables 
are involved, at most four independent four-momenta are available in the 
Lorentz frame. Therefore, the hadron tensor can be expanded on the basis of 
four independent variables~\cite{Oxford}. In the case of two-nucleon emission 
different choices are possible. One can refer to the target four-momentum 
$P^\mu$, the momentum transfer $q^\mu$, the ejectile four-momentum $p_1'^\mu$ 
and the four vector
\begin{equation}
\xi^\mu = \epsilon^{\alpha \beta \gamma \mu} q_\alpha p'_{1 \beta} P_\gamma.
\label{eq:csi}
\end{equation}
The expansion coefficients can depend on the invariants $Q^2, P \cdot q, 
q \cdot p'_1, q \cdot p'_2, P \cdot p'_1, P \cdot p'_2, p'_1 \cdot p'_2$ and 
$\xi \cdot s_1$, where $s_1$ is the spin of the outgoing nucleon with momentum 
$\p'_1$, which are scalars and $P \cdot s_1, q \cdot s_1, p'_2 \cdot s_1$ and 
$\xi \cdot p'_2$, which are pseudoscalars.
It is worthwhile to notice that both scalars and pseudoscalars, both 
dependent on and independent of $s_1$, are available.

The four-vector $s_1$, in the reference frame where $\p'_1 = 0$, is given by 
$s_1$ = (0, $\s_1$) and can be boosted in the laboratory frame as~\cite{Pick}
\begin{equation}
s_1^\mu = \left( \frac {\s_1 \cdot \p'_1} {m} , \s_1 + \frac {\s_1 \cdot \p'_1} 
{m(E'_1 + m)} \p'_1 \right).
\label{eq:s1}
\end{equation}
Therefore, we obtain the invariants
\begin{equation}
\xi \cdot s_1 = M_T \, \q \times \p'_1 \cdot \s_1,
\label{eq:csis}
\end{equation}
\begin{equation}
P \cdot s_1 = \frac {M_T} {m} \p'_1 \cdot \s_1,
\label{eq:ps1}
\end{equation}
where $m$ is the nucleon mass and $M_T$ the target mass, and 
\begin{equation}
q \cdot s_1 = \left( \frac {\omega} {m} - \frac {\q \cdot \p'_1}
{m(E'_1 + m)} \right)  \p'_1 \cdot \s_1 - \q \cdot \s_1.
\label{eq:qs1}
\end{equation}

It is clear from the above equations that $\xi \cdot s_1$ is proportional 
to $\hat{\N} \cdot \s_1$, $P \cdot s_1$ to $\hat{\L} \cdot \s_1$ and 
$q \cdot s_1$ has a component proportional to $\hat{\S} \cdot \s_1$.
Since we can obtain from these invariants both scalars and pseudoscalars 
which linearly contain the spin, il follows that no restrictions due to
parity conservation can in general be derived concerning the dependence 
of the different structure functions on the components of the spin, 
in contrast with what happens in the (${\vec e},e'{\vec N}$) reaction.

These restrictions are recovered in the particular case of a reaction 
where the outgoing nucleon momenta $\p'_1$ and $\p'_2$ and the momentum 
transfer $\q$ lie all in the same plane. In this case the invariant
$\xi \cdot p'_2$ = 0, and neither scalars proportional to $\hat{\L} \cdot \s_1$ 
and $\hat{\S} \cdot \s_1$, nor pseudoscalars proportional to 
$\hat{\N} \cdot \s_1$ are obtained. This result is independent of the electron 
plane and corresponds to the typical situation, in an unrestricted kinematics,
of the (${\vec e},e'{\vec N}$) reaction.

In general, for the (${\vec e},e'{\vec N}N$) reaction, the dependence of 
the structure functions on the spin can be explicited remembering that for 
a particle with spin $\frac {1} {2}$ only a linear dependence is allowed,
as 
\begin{equation}
f_{\lambda\lambda'} = h^{\mathrm{u}}_{\lambda\lambda'} + \hat{\s} \cdot 
\h_{\lambda\lambda'},
\label{eq:hs1}
\end{equation}    
where $\hat{\s}$ is the unit vector in the spin space. Thus, when the 
polarization of the outgoing nucleon is considered, 36 structure functions
$h^{\mathrm{u}}_{\lambda\lambda'}$ and $h^i_{\lambda\lambda'}$ are obtained. 
The explicit expressions of these structure functions in terms of the 
components of the hadron tensor $W^{\mu\nu}_{\alpha\alpha'}$ can be easily 
obtained from Table I, by simply substituting the quantities $W^{\mu\nu}$, 
wherever they appear, with the following expressions:
\begin{eqnarray}
W^{\mu\nu}_{++} + W^{\mu\nu}_{--} \,\,\,\, &  {\mathrm {for}} &  \,\,
{\mathrm{ i = u,}} \nonumber \\
W^{\mu\nu}_{+-} + W^{\mu\nu}_{-+}   \,\,\,\, & {\mathrm {for}} & \,\, 
{\mathrm{ i = x,}} \nonumber \\
\mathrm {i} (W^{\mu\nu}_{++} - W^{\mu\nu}_{--}) \,\,\,\, & {\mathrm {for}} & 
\,\, {\mathrm{ i = y,}} \nonumber \\
W^{\mu\nu}_{++} - W^{\mu\nu}_{--}  \,\,\,\, & {\mathrm {for}} & \,\, 
{\mathrm{ i = z.}}
\label{eq:wmn}
\end{eqnarray} 
Usually, the quantities $\h_{\lambda\lambda'}$ and $\h'_{\lambda\lambda'}$ 
are projected onto the basis of unit vectors $\hat{\N}$, $\hat{\L}$
and $\hat{\S}$ defined in Sec. II and the structure functions are thus given 
for the components $i = N, L, S$.

In the most general situation all the 36 structure functions do not vanish. 
When the electrons are on the $\x \z$ plane, i.e. when $\alpha$ = 0, the 
structure functions which are multiplied by $\sin\alpha$ or $\sin2\alpha$ 
in Eq.~(\ref{eq:cssec}) do not contribute. In this case only 24 structure 
functions contribute to the cross section and polarizations. When the two
outgoing nucleons and the momentum transfer lie in the same plane, we 
recover formally the typical situation of the (${\vec e},e'{\vec N}$) 
reaction~\cite{Oxford,eep} and only 18 structure functions do not vanish. In 
all these cases all the components of the polarization and of the polarization 
transfer coefficient are in principle different from zero. In a coplanar 
kinematics, i.e. when the momenta of the electrons and of the outgoing 
nucleons lie in the same plane, we recover the typical situation of the 
(${\vec e},e'{\vec N}$) reaction in coplanar kinematics, where only 12 
structure functions contribute and only $P^N, P'^L$ and $P'^S$ do not vanish.
Finally, in the super-parallel kinematics only 5 structure functions survive,  
as in the parallel kinematics of the (${\vec e},e'{\vec N}$) reaction.

%%%%%%%%%%%%%%%%%%%%%%%%%%%%%%%%%%%%%%%%

%%%%%%%%%%%%%%
%%%
%%%   TABLES
%%%
%%%%%%%%%%%%%%
\begin{table}
\begin{tabular}{|cccc|}
\hline
& & &\\
& $f_{00} = \frac{|\q|^2}{Q^2}$ \, $W^{00}$ &  $f_{11} = W^{xx} + W^{yy}$ &\\
& & &\\
& $f_{01} = - \frac{|\q|}{Q}\sqrt{2}$\, $(W^{0x} + 
W^{x0})$ &  $\overline f_{01} = \frac{|\q|}{Q}\sqrt{2}$\, $(W^{0x} + 
W^{x0})$ &\\
& & &\\
& $f_{1-1} = W^{yy} - W^{xx} $ &  $\overline f_{1-1} = W^{xy} + W^{yx}$ &\\
& & &\\
& $f'_{01} = \frac{|\q|}{Q}\, \mathrm{i} \,\sqrt{2}$\, $(W^{0x} - 
W^{x0})$ &  $\overline f'_{01} = \frac{|\q|}{Q}\, \mathrm{i} \,\sqrt{2}$\, 
$(W^{0y} - W^{y0})$ &\\
& & &\\
& $f'^{11} = - \mathrm{i} (W^{xy} - W^{yx})$ &  &\\
& & &\\
\hline
\end{tabular}

\bigskip
\caption[Table I]{
Structure functions in terms of the components of the hadron tensor.
\label{tab:ff}
}
\end{table}
%%%%%%%%%%%%%%
%%
%% FIGURES
%%
%%%%%%%%%%%%%%
%%%%%%%%%%%%%%

\begin{thebibliography}{100}

\bibitem{Gottfried} 
K.~Gottfried,                                                
\newblock Nucl. Phys.  {\bf 5}, 557 (1958); Ann. of Phys. {\bf 21}, 63
(1963).

\bibitem{Oxford}
S.~Boffi, C.~Giusti, F.~D.~Pacati and M.~Radici, 
\newblock {\em Electromagnetic Response of Atomic Nuclei}, Oxford Studies in 
Nuclear Physics (Clarendon Press, Oxford, 1996).

\bibitem{Gerco} 
C.~J.~G.~Onderwater, K.~Allaart, E.~C.~Aschenauer, Th.~S.~Bauer, D~.J.~Boersma,
E.~Cisbani, S.~Frullani, F.~Garibaldi, W.~J.~W.~Geurts, D.~Groep, 
W.~H.~A.~Hesselink, M.~Iodice, E.~Jans, N.~Kalantar-Nayestanaki, 
W.~-J.~Kasdorp, C.~Kormanyos, L.~Lapik\'{a}s, J.~J.~van Leeuwe, R.~De Leo, 
A.~Misiejuk, A.~R.~Pellegrino, R.~Perrino, R.~Starink, M.~Steenbakkers, G.~van 
der Steenhoven, J.~J.~M.~Steijger, M.~A.~van Uden, G.~M.~Urciuoli, 
L.~B.~Weinstein, and H.~W.~Willering, 
\newblock Phys. Rev. Lett. {\bf 78}, 4893 (1997).

\bibitem{NIKHEF} 
C.~J.~G.~Onderwater, K.~Allaart, E.~C.~Aschenauer, Th.~S.~Bauer, D~.J.~Boersma,
E.~Cisbani, W.~.H.~Dickhoff, S.~Frullani, F.~Garibaldi, W.~J.~W.~Geurts, 
C.~Giusti, D.~Groep, W.~H.~A.~Hesselink, M.~Iodice, E.~Jans,
N.~Kalantar-Nayestanaki, W.~-J.~Kasdorp, C.~Kormanyos, L.~Lapik\'{a}s,
J.~J.~van Leeuwe, R.~De Leo, A.~Misiejuk, H.~M\"uther, F.~D.~Pacati,
A.~R.~Pellegrino, R.~Perrino, R.~Starink, M.~Steenbakkers, G.~van der
Steenhoven, J.~J.~M.~Steijger, M.~A.~van Uden, G.~M.~Urciuoli, L.~B.~Weinstein,
and H.~W.~Willering, 
\newblock Phys. Rev. Lett. {\bf 81}, 2213 (1998).

\bibitem{Ronald} 
R.~Starink, M.~F.~Van Batenburg, E.~Cisbani, W.~.H.~Dickhoff, S.~Frullani, 
F.~Garibaldi, C.~Giusti, D.~L.~Groep, P.~Heimberg, W.~H.~A.~Hesselink, 
M.~Iodice, E.~Jans, L.~Lapik\'{a}s, R.~De Leo, C.~J.~G.~Onderwater, 
F.~D.~Pacati, R.~Perrino, J.~Ryckebusch, M.~F.~M.~Steenbakkers, J.~A.~Templon, 
G.~-M.~Urciuoli and L.~B.~Weinstein,
\newblock Phys. Lett. {\bf B}, (2000) in press.

\bibitem{Rosner}
G.~Rosner,
\newblock {\em Conference on Perspectives in Hadronic Physics},
eds. S.~Boffi, C.~Ciofi degli Atti, and M.~M.~Giannini (World Scientific,
Singapore, 1998) p.185; \\
\newblock Proceedings of the 10th Mini-Conference on Studies of
Few-Body Systems with High Duty-Factor Electron Beams, NIKHEF, Amsterdam 1999, 
p.95.

\bibitem{sf} C.~Giusti, F.~D.~Pacati, K.~Allaart, W.~J.~W.~Geurts,
W.~H.~Dickhoff and H.~M\"uther,
\newblock Phys. Rev. C {\bf 57}, 1691 (1998).

\bibitem{Gent1} J.~Ryckebusch, W.~Van Nespen and D.~Debruyne,
\newblock Phys. Lett. B {\bf 441}, 1  (1998).

\bibitem{Gent2} 
J.~Ryckebusch, D.~Debruyne and W.~Van Nespen,
\newblock Phys. Rev.  C {\bf 57}, 1319 (1998). 

\bibitem{Pick} A.~Picklesimer and J.~W.~Van Orden,
\newblock Phys. Rev. C {\bf 35}, 266 (1987).

\bibitem{GP}
C.~Giusti and F.~D.~Pacati, 
\newblock Nucl. Phys.  {\bf A535}, 573 (1991); \\
\newblock Nucl. Phys. {\bf A571}, 694 (1994).

\bibitem{eep}
C.~Giusti and F.~D.~Pacati, 
\newblock Nucl. Phys.  {\bf A504}, 373 (1989).

\bibitem{eepp}
C.~Giusti and F.~D.~Pacati, 
\newblock Nucl. Phys.  {\bf A615}, 373 (1997).

\bibitem{Wouter} W.~J.~W.~Geurts, K.~Allaart, W.~H.~Dickhoff and H.~M\"uther,
\newblock Phys. Rev. C{\bf 54}, 1144 (1996).

\bibitem{delta} 
P.\ Wilhelm, H.\ Arenh\"ovel, C.\ Giusti, and F.\ D.\ Pacati,
\newblock Z. Phys. A {\bf 359}, 467 (1997).

\bibitem{gnn} 
C.~Giusti and F.~D.~Pacati, 
\newblock Nucl. Phys.  {\bf A641}, 297 (1998).

\bibitem{delta1} 
P.~Wilhelm, J.~A.~Niskanen and  H.~Arenh\"ovel, 
\newblock Nucl. Phys.  {\bf A597}, 613 (1996).

\bibitem{Nad}
A.~Nadasen, P.~Schwandt, P.~P.~Singh, W.~W.~Jacobs, A.~D.~Bacher,
P.~T.~Debevec, M.~.D.~Kaichuk and J.~T.~Meek, 
\newblock Phys. Rev. C {\bf 23}, 1023 (1981).
 
\bibitem{pn} C.~Giusti, H.~M\"uther, F.~D.~Pacati and M.~Stauf,
\newblock Phys. Rev. C {\bf 60}, 054608 (1999).

\end{thebibliography}
\end{document}